\documentclass[12pt]{iopart}
\usepackage{iopams}
\usepackage[nosort]{cite}
\usepackage[bookmarks=false]{hyperref}
\hypersetup{colorlinks=true,citecolor=blue,linkcolor=blue,urlcolor=blue,pdfstartview=FitH,bookmarksopen=true}

\expandafter\let\csname equation*\endcsname\relax
\expandafter\let\csname endequation*\endcsname\relax
\usepackage{amsmath}
\usepackage{graphicx}

\begin{document}
\title[]{Ground-state properties of anyons in a one-dimensional lattice}
\author{Guixin Tang$^1,^2$\footnote{Corresponding author: tangguixin@hit.edu.cn}, Sebastian Eggert$^2$ and Axel Pelster$^2$}
\address{$^1$Department of Physics, Harbin Institute of Technology, 150001 Harbin, China}
\address{$^2$Physics Department and Research Center OPTIMAS, University of Kaiserslautern, 67663 Kaiserslautern, Germany}
\ead{tangguixin@hit.edu.cn}

\begin{abstract}
Using the Anyon-Hubbard Hamiltonian, we analyze the ground-state properties of anyons in a 
one-dimensional lattice. To this end we map the hopping dynamics of correlated anyons to an 
occupation-dependent hopping Bose-Hubbard model using the fractional Jordan-Wigner transformation. 
In particular, we calculate the quasi-momentum distribution of anyons, which interpolates 
between Bose-Einstein and Fermi-Dirac statistics. Analytically, we apply a modified Gutzwiller 
mean-field approach, which goes beyond a classical one by including the influence of the fractional phase of 
anyons within the many-body wavefunction. Numerically, we use the density-matrix renormalization group 
by relying on the ansatz of matrix product states. 
As a result it turns out that the anyonic quasi-momentum distribution reveals 
both a peak-shift and an asymmetry which mainly originates from the nonlocal string property. 
In addition, we determine the corresponding quasi-momentum distribution of the Jordan-Wigner transformed bosons, 
where, in contrast to the hard-core case, we also observe an asymmetry for the soft-core case, 
which strongly depends on the particle number density.
\end{abstract}



\submitto{\NJP}


\section{Introduction}

A fundamental principle of quantum statistical mechanics in three dimensions is the existence of 
two types of particles: bosons obeying Bose-Einstein statistics and fermions obeying Fermi-Dirac statistics. 
However, in a two-dimensional electron liquid, quasi-particles made of electrons in the fractional 
quantum Hall effect are charged anyons obeying fractional statistics \cite{anyon1,anyon2,anyon3,Laughlin,Wen}. 
While exchanging two anyons, the many-body wavefunction acquires a fractional phase 
$e^{\mathrm{i} \theta}$, where the statistical parameter $\theta$ varies with the interval $0<\theta<\pi$ 
and corresponds to a fractional statistics. 
As compared to bosons and fermions, anyons exhibit a wide range of previously unexpected properties 
and the concept of anyons plays an important role in numerous studies of condensed matter physics 
and of topological quantum computation \cite{TQC1,TQC2,TQC3,TQC4,TQC5,TQC6,TQC7,TQC8}. 
In order to generalize the Bose-Einstein and Fermi-Dirac statistics by allowing a maximal finite integer 
particle number occupying the same quantum state, Gentile supplied an intermediate statistics that had 
been proven to be also valid for a $q$-fermion \cite{Gentile1,Gentile2,Dutt,Biedenharn,Macfarlane,Parthasarathy}. 
However, Y. Shen et al. pointed out that the anyon statistics was not a complete analogue 
of the Gentile statistics \cite{Long}. Based on a generalized Pauli exclusion principle, 
Haldane provided a useful concept of fractional statistics in arbitrary dimensions \cite{Haldane}. 
Polychronakos also suggested another form of the fractional exclusion statistics \cite{Polychronakos}. 
However, it was shown via the virial expansion that all these fractional statistics proposed 
in the literature do not apply for anyons \cite{Long, Avinash}. 

In one dimension (1D), anyons were realized as low-energy elementary excitations of the Hubbard model 
of fermions with correlated hopping processes \cite{bond-charge}. 
Alternatively, it was suggested to create anyons by bosons with complex-valued occupation-dependent hopping 
amplitudes by photon-assisted tunneling in 1D optical lattices \cite{Marco,Ruichao}.
Recently, S. Greschner and L. Santos proposed a Raman scheme to improve the proposal of T. Keilmann et al. 
in Ref. \cite{Marco} and deduced a rich ground-state physics including Mott-insulators 
with attractive interactions, pair-superfluids, dimer phases, and multi-critical points \cite{Greschner}. 
An even simpler scheme for realizing the physics of 1D anyons with ultracold bosons in an optical lattice 
has recently been proposed in Ref. \cite{AndrePoster}. 
It relies on lattice-shaking-induced resonant tunneling against the energy off-sets created by the combination of 
both a potential tilt and on-site repulsion. In contrast to the above mentioned former proposals based on 
internal atomic degrees of freedom, no lasers additional to those already used for the creation of 
the optical lattice are required \cite{AndrePoster}.

The physical properties of 1D anyons are intriguing and complicated. Theoretically, 
using the generalized coordinate Bethe ansatz method, Kundu obtained the exact solution 
of a 1D anyon gas with the delta-function potential, which shows that the effective interaction 
among the anyons is modified by the anyonic statistical parameter \cite{Kundu}. 
Furthermore, the 1D anyon gas with a special interaction potential has been investigated 
in various 1D systems in order to get the ground-state properties of the one-dimensional anyon gas 
\cite{Batchelor1,Batchelor2,Batchelor3,Schen2008,Santos,Calabrese}. 
For anyons in the 1D lattice systems one investigated, for instance, the statistically induced 
phase transition \cite{Marco}, quantum scaling properties \cite{bond-charge}, 
dynamics properties of hard-core anyons \cite{SChen2012,Wright}, quantum walks \cite{Limin,Preiss}, 
and anyonic Bloch oscillations \cite{Stefano}. 
For the hard-core anyons in the 1D optical lattice, Y. Hao et al. investigated 
how the fractional statistics affected the ground-state properties by mapping the hard-core anyonic Hamiltonian 
to a noninteracting fermionic system with a generalized Jordan-Wigner transformation \cite{SChen2009}. 
By calculating the one-particle Green’s function of the ground state, 
the reduced one-body density matrix and, thus, the quasi-momentum distributions can be obtained 
for different statistical parameters of anyons. 
The results showed that the momentum distributions in the Bose and Fermi limit turned out to be symmetric, 
but those of anyons were in general asymmetric and shifted due to the fractional statistics that anyons obey. 

In this paper we complement the previous studies of the ground-state properties of a 1D quantum gas 
of anyons confined in optical lattices. To this end we map the Anyon-Hubbard model 
to an occupation-dependent hopping Bose-Hubbard model with the help of a fractional version of
the Jordan-Wigner transformation. This mapping has the consequence that the Hilbert space of anyons 
can be constructed from that of bosons, so that one has access to the two-point correlation function 
of either the original anyonic or the Jordan-Wigner transformed bosonic creation and annihilation operators.
With this we investigate the quasi-momentum distributions of either bosons or anyons, 
the latter interpolating between Bose-Einstein statistics and Fermi-Dirac statistics. 
Firstly, in the hard-core limit, we determine the quasi-momentum distributions of anyons 
with density-matrix renormalization group calculations, which numerically reproduces the results 
of Y. Hao et al. \cite{SChen2009}. In addition, by suggesting a modified Gutzwiller mean-field approach 
to include the influence of the fractional phase of anyons within the many-body wavefunction, 
we obtain an approximative analytic expression for the numerical results of Ref. \cite{SChen2009} 
both for a finite system and in the thermodynamic limit. 
In particular, we analyze in detail how peak-shift and asymmetry of the quasi-momentum distribution 
of anyons depend on both the fractional phase $\theta$ and the particle number density $n_0$.
Furthermore, we extend the findings of Ref. \cite{SChen2009} by working out also the more general soft-core case. 
The quasi-momentum distribution of anyons reveals the pseudofermion property 
at the fractional phase  $\theta=\pi$. Surprisingly, the quasi-momentum distributions of 
the Jordan-Wigner transformed bosons shows a density-dependent asymmetry 
which is not found for the hard-core case.

The outline of this paper is as follows: in Sec. \ref{AnyonHubbardModel} the Anyon-Hubbard model 
and the mapping between anyons and bosons are discussed. 
Both the classical and the modified Gutzwiller mean-field approach is introduced in Sec. \ref{GW-MF}. 
The ground-state properties of the 1D Anyon-Hubbard model are determined by studying 
the quasi-momentum distribution of anyons and bosons in both the hard-core and 
the soft-core case in Sec. \ref{hard-core} and Sec. \ref{soft-core} respectively. 
Conclusions are given in Section \ref{Conclusions}. 

\section{Anyon-Hubbard model}
\label{AnyonHubbardModel}

The hopping dynamics of correlated anyons on a 1D lattice is described by 
the Anyon-Hubbard (AH) Hamiltonian \cite{Marco}
\begin{equation}\label{AHM}
\hat{H}^{a}=-J\sum_{j=1}^{L} (\hat{a}_{j}^{\dagger}\hat{a}_{j+1}+\mathrm{h.c.})
           +\frac{U}{2}\sum_{j=1}^{L} \hat{n}_{j}(\hat{n}_{j}-1),
\end{equation}
where $J>0$ denotes the tunneling amplitude connecting two neighbouring sites, 
$U$ stands for the on-site interaction energy, 
$\hat{n}_{j}=\hat{a}_{j}^{\dagger}\hat{a}_{j}$ represents the number operator at site $j$, and 
the operators $\hat{a}_{j}^{\dagger}$, $\hat{a}_{j}$ create or annihilate an anyon on site $j$. 
For 1D anyons, the operators $\hat{a}_{j}^{\dagger}$ and $\hat{a}_{j}$  
obey the generalized commutation relations \cite{Marco,Kundu}
\begin{equation}\label{CCR}
\begin{split}
\hat{a}_{j}\hat{a}_{k}^{\dagger}-e^{-\mathrm{i}\theta\mathrm{sgn}(j-k)}\hat{a}_{k}^{\dagger}\hat{a}_{j}
&=\delta_{jk} \, , \\
\hat{a}_{j}\hat{a}_{k}-e^{\mathrm{i}\theta\mathrm{sgn}(j-k)}\hat{a}_{k}\hat{a}_{j} &= 0 \, ,
\end{split}
\end{equation}
where $\theta$ is the statistical exchange phase, the sign function $\mathrm{sgn}(j-k)=-1, 0, +1$ 
for $j<k$, $j=k$, and $j>k$, respectively. 
Note that, since the sign function $\mathrm{sgn}(j-k)=0$ for $j=k$, 
two particles on the same site behave as ordinary bosons irrespective of the statistical parameter $\theta$. 
Moreover, anyons with the statistical exchange phase $\theta=\pi$ are pseudofermions, i.e.  
while being bosons on-site, they are fermions off-site. 

Here we use an exact Anyon-Boson mapping in 1D in terms of a fractional version of 
the Jordan-Wigner transformation \cite{Marco}
\begin{equation}\label{mapping}
\hat{a}_{j}=\hat{b}_{j}\exp\left(\mathrm{i}\theta\sum_{i=1}^{j-1}\hat{n}_{i}\right) \,,
\end{equation}
where the number operator reads 
$\hat{n}_{i}=\hat{a}_{i}^{\dagger}\hat{a}_{i}=\hat{b}_{i}^{\dagger}\hat{b}_{i}$ 
and $\hat{b}_{i}$, $\hat{b}_{i}^{\dagger}$ are bosons operators following the commutation relation 
$[\hat{b}_{i},\hat{b}_{j}^{\dagger}]=\delta_{ij}$ and 
$[\hat{b}_{i}^{\dagger},\hat{b}_{j}^{\dagger}]=0=[\hat{b}_{i},\hat{b}_{j}]$. 
Inserting the Anyon-Boson mapping (\ref{mapping}) into Eq. (\ref{AHM}), 
the Hamiltonian $\hat{H}^{a}$ can be rewritten as \cite{Marco} 
\begin{equation}\label{bosons}
\hat{H}^{b}=-J\sum_{j=1}^{L} (\hat{b}_{j}^{\dagger}\hat{b}_{j+1}e^{\mathrm{i}\theta \hat{n}_{j}}+\mathrm{h.c.})
           +\frac{U}{2}\sum_{j=1}^{L} \hat{n}_{j}(\hat{n}_{j}-1), 
\end{equation}
where the conditional hopping of bosons from right to left, i.e. $j+1 \rightarrow j$, 
occurs with an occupation-dependent amplitude $J e^{\mathrm{i}\theta \hat{n}_{j}}$. 
If the target site $j$ is unoccupied, the hopping amplitude is simply $J$. 
If it is occupied by one boson, the amplitude becomes complex and reads $Je^{\mathrm{i}\theta}$, and so on. 
We emphasize that the hard-core limit of anyons in Eq. (\ref{AHM}) coincides with that of bosons 
in Eq. (\ref{bosons}) due to $\hat{n}_{i}=\hat{a}_{i}^{\dagger}\hat{a}_{i}=\hat{b}_{i}^{\dagger}\hat{b}_{i}$. 
It is also clear that the mapped bosonic Hamiltonian (\ref{bosons}) describes local occupation numbers 
beyond the hard-core limit $n_j > 1$. 

Using the nonlocal exact mapping (\ref{mapping}) between anyons and bosons, the Anyon-Hubbard Hamiltonian 
(\ref{AHM}) leads to a bosonic Hamiltonian Eq. (\ref{bosons}) that can be solved either analytically or 
numerically in order to determine the ground-state properties of anyons in 1D lattice systems. 
Theoretically, since the reflection parity symmetry in the Hamiltonian is broken, we suggest below a modified 
Gutzwiller mean-field, which goes beyond the classical one found in literature \cite{Rokhsar,Krauth,Jaksch,Schroll} 
in order to include the influence of the fractional phase of anyons on the hopping dynamics. 
To get numerical results, we use the density-matrix renormalization group (DMRG) \cite{DMRG01,DMRG02,DMRG1,DMRG2,DMRG3}, 
which was already applied to the realm of anyons, for instance, in Ref. \cite{Marco}. 
To this end, we rely on the ansatz of matrix product states (MPS) with the system length $L$ and 
open boundary conditions, which is more efficient than periodic boundaries \cite{MPS1,MPS2,MPS3,MPS4,MPS5}. 
The code is based on a variational ansatz using MPS with the restricted subspace of integer filling, 
where our simulations admit a maximum of five particles per site 
and the maximum bond dimension of MPS equals to 1000 \cite{MPS6}. 
Moreover, Refs. \cite{MPS3,MPS4,MPS6} show the respective details how to calculate 
the expectation of the string operator in Eq. (\ref{mapping}) and 
the correlation functions in DMRG with MPS ansatz.

\section{Mean-field approximation}
\label{GW-MF}

At first we work out a Gutzwiller mean-field approach, which turns out to provide qualitative satisfactory 
results for the quasi-momentum distributions. In a Gutzwiller (GW) approach for bosons, 
the many-particle state $|G\rangle$ is generically approximated by a product state of 
single lattice-site states $|\Phi_{j}\rangle$, 
which can be expressed as a superposition of different number states on a lattice site \cite{Schroll} 
\begin{equation}\label{Awf_GW}
|G\rangle = \prod_{j} |\Phi_{j}\rangle = 
\prod_{j} \left( \sum_{n=0}^{n_{\mathrm{max}}} f_n^{(j)} |n\rangle \right) \, .
\end{equation}
Here $n_{\mathrm{max}}$ represents a truncation at some sufficiently large maximal particle number and 
$f_n^{(j)}$ denotes the Gutzwiller probability amplitude of finding $n$ bosons on site $j$, 
which is normalized such that 
\begin{equation}\label{normalization}
\sum_{n=0}^{n_{\mathrm{max}}} |f_n^{(j)}|^2 = 1  \, .
\end{equation} 

The total energy of the system 
\begin{equation}\label{tot_ene}
\mathcal{E}_{\mathrm{tot}} = \mathcal{E}_{\mathrm{int}} + \mathcal{E}_{\mu} + \mathcal{E}_{\mathrm{kin}} 
\end{equation}
then follows from determining the ground-state expectation values of the respective operators 
according to the Appendix. As a consequence, the interaction energy (\ref{E_int}), 
the chemical potential term (\ref{E_mu}), 
and the kinetic energy (\ref{E_kin}) depend on the Gutzwiller coefficients $f_n^{(j)}$. 
Due to the polar decomposition $f_n^{(j)} = F_n^{(j)} e^{ \mathrm{i}\alpha_n^{(j)} }$, 
the total energy (\ref{tot_ene}) has to be minimized with respect to both the absolute values $F_n^{(j)}$ 
and the phases $\alpha_n^{(j)}$. 

In the classical GW mean-field approach of a homogeneous system, to which we refer as cGW in the following, 
one assumes that the ground state is a product of identical states 
on each of the $L$ lattice sites \cite{Rokhsar,Krauth,Jaksch,Schroll}, i.e., 
\begin{equation}
|\Phi_{j}\rangle = |\Phi \rangle , \quad f_n^{(j)} = f_n \,.
\end{equation}
This has the consequence 
\begin{equation}
F_n^{(j)} = F_n, \quad \alpha_n^{(j)} = \alpha_n \,,
\end{equation}
i.e. one has the identical wave function on each lattice site. 

Due to the occupation-dependent amplitude $Je^{\mathrm{i}\theta \hat{n}_j}$ 
for nearest-neighbour hopping in the bosonic Hamiltonian (\ref{bosons}), 
the reflection parity symmetry is broken. Therefore, we suggest to drop the cGW constraint 
of having identical states on each site to a modified GW mean-field approach (mGW). 
To this end, we note that the kinetic energy Eq. (\ref{E_kin_f}) in the Appendix 
depends on the difference 
$\beta_n^{(j)} \equiv \alpha_n^{(j)} - \alpha_{n+1}^{(j)}$ of the phases of the probability amplitudes 
$f_n^{(j)}$ and $f_{n+1}^{(j)}$ of finding $n$ and $n+1$ bosons on site $j$, respectively. 
If all phase differences would be identical, we would recover cGW. 
Therefore, we define mGW according to 
\begin{equation}\label{f_mGW}
F_n^{(j)} = F_n \,, \quad  \Delta\beta_n^{(j)} = \beta_n^{(j)} - \beta_{n}^{(j+1)} \equiv \Delta\beta_n \,.
\end{equation}
In order to fix the values of the phase differences $\beta_n^{(j)}$, 
we complement (\ref{f_mGW}) by the additional assumption 
\begin{equation}
\beta_n^{(j)} + \beta_{n}^{(j+1)} \equiv 2\beta_{n} \,.
\end{equation}
This means that the absolute value $F_n^{(j)}$ of the probability amplitude $f_n^{(j)}$ is identical 
on each site and the non-vanishing phase difference $\Delta\beta_n^{(j)}$ 
between two nearest-neighbour sites is the same in the whole chain. 

The application of cGW and mGW is performed in the Appendix for both the hard-core case in Sec. \ref{appen:HC}, 
which means $n_{\mathrm{max}}=1$, and the soft-core case by assuming $n_{\mathrm{max}}=2$ in Sec. \ref{appen:SC}. 
Provided that Gutzwiller amplitudes and phases have been determined, one can calculate 
the two-point correlation function of bosons, which has the rather simple expression 
\begin{equation}\label{bibj}
\langle \hat{b}_i^{\dagger} \hat{b}_j \rangle = \delta_{ij} \langle \hat{n}_i \rangle 
   + (1-\delta_{ij}) \langle \hat{b}_i^{\dagger} \rangle \langle \hat{b}_j \rangle \,.
\end{equation}
The correlation function of anyons can be written as 
\begin{equation}\label{aiaj_src}
\langle \hat{a}_i^{\dagger} \hat{a}_j \rangle = 
\delta_{ij} \langle \hat{n}_i \rangle +(1-\delta_{ij})\langle \hat{a}_i^{\dagger} \hat{a}_j \rangle \, .
\end{equation}
For the case $i<j$ and $i>j$ we get correspondingly, 
\begin{subequations}\label{aiaj}
\begin{align}
\langle \hat{a}_i^{\dagger} \hat{a}_j \rangle 
&\xrightarrow[]{i<j}  \langle \hat{b}_{i}^{\dagger} e^{\mathrm{i}\theta\hat{n}_i} \rangle 
   \bigl( \prod_{i<l<j} \langle e^{\mathrm{i}\theta\hat{n}_{l}} \rangle \bigr)
   \langle \hat{b}_j \rangle \, ,\label{aiaj1} \\
\langle \hat{a}_i^{\dagger} \hat{a}_j \rangle 
&\xrightarrow[]{i>j} \langle e^{-\mathrm{i}\theta\hat{n}_j} \hat{b}_j \rangle  
   \bigl( \prod_{j<l<i} \langle e^{-\mathrm{i}\theta\hat{n}_{l}} \rangle \bigr)
   \langle \hat{b}_{i}^{\dagger} \rangle \, , \label{aiaj2}
\end{align}
\end{subequations}
where the respective expectation values depend on the Gutzwiller amplitudes $f_n^{(j)}$ 
according to Eq. (\ref{op_aver}) in the Appendix. 

Furthermore, in order to investigate the effect of the anyon statistical parameter $\theta$, 
we will focus on density distributions of bosons and anyons in the quasi-momentum space, 
which are believed to be interesting \cite{Marco} 
and are defined via the Fourier transformation of the correlation function 
\begin{subequations}\label{nk}
\begin{align}
\langle \hat{n}_k^{(\mathrm{b})} \rangle  &= \frac{1}{L} \sum_{ij} e^{\mathrm{i} k (x_i-x_j)} 
                                           \langle \hat{b}_i^{\dagger} \hat{b}_j \rangle \, ,\label{nk_b} \\
\langle \hat{n}_k^{(\mathrm{a})} \rangle  &= \frac{1}{L} \sum_{ij} e^{\mathrm{i} k (x_i-x_j)} 
                                           \langle \hat{a}_i^{\dagger} \hat{a}_j \rangle \, .\label{nk_a}
\end{align}
\end{subequations}

In the following, we are going to find the ground state of a homogeneous system for a given particle number 
density $n_0>0$, ratio $J/U$, and varying statistical parameter $\theta$. 
Thus, combining the analytic approaches cGW, mGW with numerical results from DMRG, we determine the correlation 
functions and, thus, investigate the quasi-momentum distributions of the ground state. 

\section{Hard-core anyons}
\label{hard-core}

In order to illustrate the applicability of the Gutzwiller mean-field theory, we investigate at first 
the simplest case of hard-core anyons. This means that we assume $U/J\to\infty$, 
where bosons are impenetrable and each site contains at most one particle, i.e., $n_{\mathrm{max}}=1$. 
In the following, we present only the calculations of the correlation function 
and the quasi-momentum distribution, while the mathematical details for minimizing 
the total energy are relegated to Sec. \ref{appen:HC} of the Appendix. 

\begin{figure}[t]
  \centering
    \includegraphics[width=0.8\textwidth]{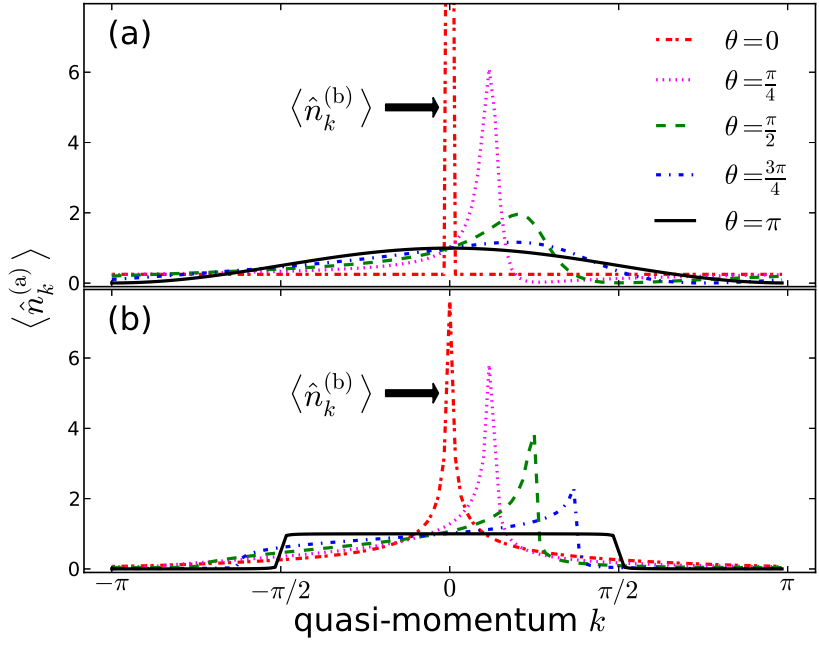}
    \caption{Quasi-momentum distribution of the ground state 
    in a hard-core system ($n_0=N/L=0.5$, $L=120$). 
    (a) Gutzwiller approach and (b) DMRG for anyons and bosons.
    }
\label{fig:nk_L120N60_hc}
\end{figure}

According to Eq. (\ref{bibj}), the correlation function of bosons reads 
\begin{equation}\label{bibj_hc}
\langle \hat{b}_i^{\dagger} \hat{b}_j \rangle 
= \delta_{ij} n_0 + (1-\delta_{ij}) n_0 (1-n_0) e^{\mathrm{i} (\beta_0^{(i)} - \beta_0^{(j)}) } \, ,
\end{equation}
which reduces to 
\begin{equation}\label{bibj_hc1}
\langle \hat{b}_i^{\dagger} \hat{b}_j \rangle = \delta_{ij} n_0^2 + n_0 (1-n_0) \, ,
\end{equation}
since $\Delta\beta_0 = \beta_0^{(j)} - \beta_0^{(j+1)} = 2m\pi$ (see Sec. \ref{appen:HC}). 
Using Eq. (\ref{nk_b}), the ground-state quasi-momentum distribution of bosons can be written as 
\begin{equation}\label{HC:nk_b_hc}
\langle \hat{n}_k^{(\mathrm{b})} \rangle  = n_0^2 + n_0 (1-n_0) L \delta_{k,0} \, ,
\end{equation}
which is $\theta$-independent and the number of particles in the condensate is $N_0 = N(1-n_0)$. 
Fig. \ref{fig:nk_L120N60_hc} shows the resulting ground-state quasi-momentum distribution 
in a hard-core system at density $n_0=0.5$. 

Since $\langle \hat{n}_k^{(\mathrm{b})} \rangle$ in Eq. (\ref{HC:nk_b_hc}) is $\theta$-independent 
in the GW approach, if one changes $\theta$ from $0$ to $\pi$, all quasi-momentum distributions 
with different $\theta$ coincide as shown in Fig. \ref{fig:nk_L120N60_hc} (a). 
The quasi-momentum distribution of bosons $\langle \hat{n}_k^{(\mathrm{b})} \rangle$ in DMRG as shown 
in Fig. \ref{fig:nk_L120N60_hc}(b) turns out to be also $\theta$-independent 
because the phase factor disappears in the bosonic Hamiltonian (\ref{bosons}) in the hard-core limit 
and the systems reduces to a hard-core Bose-Hubbard model. 
As a consequence, bosons condense at zero quasi-momentum in both the GW approach and DMRG. 
However, the bosonic population $N(1-n_0)$ of $k=0$ in the GW approach is much larger than that in DMRG 
because the mean-field approach does not contain quantum fluctuations 
which generally broaden the quasi-momentum distributions. 

In the Bose limit $\theta = 0$, we have $\hat{a}_j = \hat{b}_j$ and 
$\langle \hat{a}_i^{\dagger} \hat{a}_j \rangle = \langle \hat{b}_i^{\dagger} \hat{b}_j \rangle$, thus 
\begin{equation}\label{nk_a_hc0_L}
\langle\hat{n}_k^{(\mathrm{a})} (\theta = 0)\rangle=\langle\hat{n}_k^{(\mathrm{b})}\rangle \, .
\end{equation}
which is shown in Fig. \ref{fig:nk_L120N60_hc}(a) and (b) for the density $n_0=0.5$. 

According to Eqs. (\ref{aiaj}) and (\ref{en_aver}) in the Appendix, 
the correlation function of anyons reduces to 
\begin{subequations}\label{aiaj_hc}
\begin{align}
\langle \hat{a}_i^{\dagger} \hat{a}_j \rangle  & \xrightarrow[]{i<j} 
     n_0 (1-n_0) \prod_{i<l<j} \bigl[ (1-n_0)+n_0 e^{\mathrm{i}\theta} \bigr] \, ,\label{aiaj1_hc} \\
\langle \hat{a}_i^{\dagger} \hat{a}_j \rangle  & \xrightarrow[]{i>j} 
     n_0 (1-n_0) \prod_{j<l<i} \bigl[ (1-n_0)+n_0 e^{-\mathrm{i}\theta} \bigr] \, . \label{aiaj2_hc}
\end{align}
\end{subequations}
Using Eq. (\ref{nk_a}), the ground-state quasi-momentum distribution of anyons 
can be written in the $\theta$-dependent form for a finite open system 
\begin{equation}\label{nk_a_hc}
\langle \hat{n}_k^{(\mathrm{a})} \rangle  
 = n_0 + n_0 (1-n_0) \frac{1}{L} \Bigl\lbrace (L-1)(e^{-\mathrm{i}k} + e^{\mathrm{i}k}) \! +\! \sum_{m=2}^{L-1} (L\!-\!m) \Bigl[ e^{-\mathrm{i}km} z^{m-1} 
       \! +\! e^{\mathrm{i}km} z^{*m-1} \Bigr] \Bigr\rbrace \, ,
\end{equation}
where $z=re^{\mathrm{i}\varphi}=x+\mathrm{i}y \equiv (1-n_0)+n_0e^{\mathrm{i}\theta}$, 
thus $r=1$ if $\theta = 0$ or $n_0 = 1$, otherwise $r<1$. 

In the Fermi limit $\theta = \pi$, we obtain from (\ref{nk_a_hc}) 
\begin{equation}\label{nk_a_hc1_L}
\langle \hat{n}_k^{(\mathrm{f})} \rangle 
 = n_0 + 2n_0 (1-n_0) \Bigl\{ \frac{\cos(k)-x}{1-2x\cos(k)+x^2} 
 - \frac{(1-x^L)[\cos(k)(1+x^2)-2x]}{L[1-2x\cos(k)+x^2]} \Bigl\} \, ,
\end{equation}
where $\langle \hat{n}_k^{(\mathrm{f})} \rangle \equiv \langle \hat{n}_k^{(\mathrm{a})} (\theta=\pi) \rangle$, 
$x\equiv 1-2n_0 \in (-1,+1)$ if $0<n_0<1$. In the thermodynamic limit, this reduces to 
\begin{equation}\label{nk_a_hc1}
\langle \hat{n}_k^{(\mathrm{f})} \rangle 
\xrightarrow[|x|<1]{L\rightarrow \infty} n_0 + 2n_0 (1-n_0) \frac{\cos(k)-x}{1-2x\cos(k)+x^2} \, .
\end{equation}
For the system at $n_0=0.5$, the GW approach gives  
\begin{equation}\label{nk_a_hc1_1}
\begin{split}
\langle \hat{n}_k^{(\mathrm{f})} \rangle 
& = 0.5 + 0.5 \bigl(1-\frac{1}{L}\bigr) \cos(k) \\
& \mspace{-18.0mu} \xrightarrow[]{L \rightarrow \infty} \langle \hat{n}_k^{(\mathrm{f})} \rangle = 0.5 + 0.5\cos(k) \, ,
\end{split}
\end{equation}
which is shown as the black solid line in Fig. \ref{fig:nk_L120N60_hc}(a). 
However, in Fig. \ref{fig:nk_L120N60_hc}(b) from DMRG, 
the quasi-momentum distribution of anyons in the Fermi limit represents a step-like distribution, 
which is the characteristic feature of free fermions. This shows that DMRG can grasp 
the fermion-like feature of anyons with statistics $\theta = \pi$, whereas the GW approach 
achieves this only approximately. 

If $K_{\mathrm{max}}^{(\mathrm{a,b)}}$ defines the quasi-momentum, 
where $\langle \hat{n}_k^{(\mathrm{a,b})} \rangle$ has its maximum, 
we have $K_{\mathrm{max}}^{(\mathrm{a})}(\theta = 0) = 0$ 
and $K_{\mathrm{max}}^{(\mathrm{a})} (\theta = \pi) = 0$, 
i.e., both the quasi-momentum distribution of the Bose and of the Fermi limit are symmetric 
about zero quasi-momentum. Indeed this results from $\delta_{k,0}$ in Eqs. (\ref{HC:nk_b_hc}) 
and (\ref{nk_a_hc0_L}) in the Bose limit and from calculating 
$\partial \langle \hat{n}_k^{(\mathrm{f})} \rangle / \partial k = 0$ 
of Eqs. (\ref{nk_a_hc1}) and (\ref{nk_a_hc1_1}) in the Fermi limit. 

In the fractional phase interpolating between the Bose and Fermi limit $0 < \theta < \pi$, 
we find from (\ref{nk_a_hc}) 
\begin{equation}\label{nk_a_hc2_L}
\langle \hat{n}_k^{(\mathrm{a})} \rangle = n_0 + \frac{n_0 (1 - n_0)}{L} 
\Bigl[\frac{(L\!-\!1)\!-\! L(ze^{-\mathrm{i}k}) \!+\! (ze^{-\mathrm{i}k})^L}
                         {e^{\mathrm{i}k} (1-ze^{-\mathrm{i}k})^2} \! + \! \mathrm{c.c.} \Bigr] \,.
\end{equation}
In the thermodynamic limit, this reduces to 
\begin{equation}\label{nk_a_hc2}
\langle \hat{n}_k^{(\mathrm{a})} \rangle  \xrightarrow[r<1]{L\rightarrow \infty} n_0\! +\! 2n_0 (1\!-\! n_0) 
    \frac{\cos(k)-r\cos(\varphi)}{1-2r\cos(k\! -\! \varphi) \! + \! r^2} ,
\end{equation}
where $r=|1-n_0+n_0e^{\mathrm{i}\theta}|$ and $\varphi = \mathrm{arg}(1-n_0+n_0e^{\mathrm{i}\theta})$. 
Fig. \ref{fig:nk_L120N60_hc} clearly show that: 
i) the respective peak of the quasi-momentum distribution is shifted to positive momentum, 
ii) the quasi-momentum distribution is asymmetric about its own peak, and 
iii) the quasi-momentum distributions broaden and flatten with increasing $\theta$. 
Comparing Fig. \ref{fig:nk_L120N60_hc} (a) with (b) or 
$\langle \hat{b}_i^{\dagger} \hat{b}_j \rangle$ in Eq. (\ref{bibj_hc1}) with 
$\langle \hat{a}_i^{\dagger} \hat{a}_j \rangle$ in Eq. (\ref{aiaj_hc}), 
we conclude that the nonlocal string property of anyons, which is denoted by 
$\prod_{i<l<j} \bigl[ (1-n_0)+n_0 e^{\mathrm{i}\theta} \bigr]$ or 
$\prod_{j<l<i} \bigl[ (1-n_0)+n_0 e^{-\mathrm{i}\theta} \bigr]$ in Eq. (\ref{aiaj_hc}), 
is the reason why anyons prefer to appear on positive momentum. 
This is explained numerically in Ref. \cite{SChen2009}. 

\begin{figure}[t]
  \centering
  \includegraphics[width=0.8\textwidth]{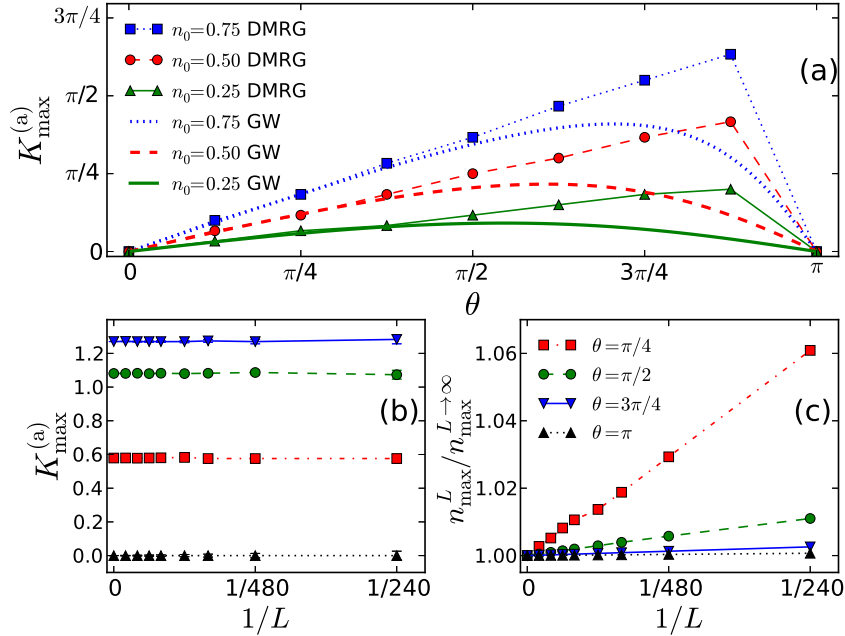}
  \caption{  
  (a) Maximum $K_{\mathrm{max}}^{(\mathrm{a)}}$ of quasi-momentum $\langle \hat{n}_k^{(\mathrm{a})} \rangle$ 
      plotted against the fractional phase $\theta$ of anyons at density $n_0$ resulting from DMRG ($L=120$) 
      and GW in the thermodynamic limit $L \rightarrow \infty$. 
  (b) $K_{\mathrm{max}}^{(\mathrm{a)}}$ plotted against the inverse system size 
      at density $n_0 = 0.75$ resulting from GW. 
  (c) Maximum $\langle \hat{n}_k^{(\mathrm{a})} \rangle$ of the finite system size $L$ divided by 
      that of the infinite system size $L \rightarrow \infty$ plotted against the inverse system size 
      at density $n_0 = 0.75$ resulting from GW.
  }
\label{fig:kmax_theta_hc}
\end{figure}

In order to find the maximum $K_{\mathrm{max}}^{(\mathrm{a)}}$ of the quasi-momentum distribution, we analytically 
evaluate $\partial \langle \hat{n}_k^{(\mathrm{a})} \rangle / \partial k = 0$ for Eq. (\ref{nk_a_hc2}), which gives 
\begin{equation}\label{Kmax}
K_{\mathrm{max}}^{(\mathrm{a)}} \!\! =\! \left\{ 
\begin{aligned}
\! \pi \!\!-\! \mathcal{K},&\,\, n_0\geq 2(\sqrt{2}\!-\!1)\text{ and }\vartheta_1 \!\leq \theta \!\leq \vartheta_2,\\
\mathcal{K},               &\,\, \text{Otherwise},
\end{aligned}\right.
\end{equation}
where we have introduced the abbreviations 
\begin{gather}
\vartheta_1 = \frac{\pi}{4} + \arcsin \bigl( \frac{2-n_0}{\sqrt{2}n_0} \bigr), \,\,\, 
\vartheta_2 = \frac{3\pi}{2} - \vartheta_1 \,,   \label{Kmax_1} \\
\mathcal{K}=\arcsin \frac{\sin\theta(2n_0-n_0^2)+n_0^2\sin\theta\cos\theta}
                         {(1-\cos\theta)(1-n_0)^2+(1+\cos\theta)}  \label{Kmax_2}  \,. 
\end{gather}

Fig. \ref{fig:kmax_theta_hc}(a) shows the resulting fractional phase $\theta$-dependence of anyonic 
$K_{\mathrm{max}}^{(\mathrm{a)}}$ at density $n_0 = 0.75$, $0.5$, and $0.25$. 
The DMRG results ($L=120$) clearly show that the peaks of the quasi-momentum distributions increase linearly from 
$K_{\mathrm{max}}^{(\mathrm{a)}}(\theta=0)=0$ (bosons) and 
come back to $K_{\mathrm{max}}^{(\mathrm{a)}}(\theta=\pi)=0$ (fermions) 
while increasing $\theta$. The GW approach shows the same linearly increasing property 
at small $\theta$, i.e. near bosons. This can be quantitatively understood by taking the limit 
$\theta \rightarrow 0$ in Eqs. (\ref{Kmax}) and (\ref{Kmax_2}):
\begin{equation}
K_{\mathrm{max}}^{(\mathrm{a)}}(\theta \rightarrow 0) \simeq \arcsin(n_0 \theta) \simeq n_0 \theta \, .
\end{equation}
At the Fermi side $\theta \rightarrow \pi$, we have correspondingly 
\begin{equation}
K_{\mathrm{max}}^{(\mathrm{a)}}(\theta \rightarrow \pi) \simeq \arcsin \frac{n_0(\pi-\theta)}{1-n_0} \simeq \frac{n_0(\pi-\theta)}{1-n_0} \, ,
\end{equation}
where the slope of $K_{\mathrm{max}}^{(\mathrm{a)}}$ from GW is smaller than that from DMRG. In general, 
the GW approach always yields a smaller shift of positive momentum in comparison with the DMRG results 
for any statistical parameter of anyons between the Fermi and the Bose case. Moreover, we find that 
both DMRG and GW give a larger peak shift for a bigger particle number density $n_0$. 

In order to determine the thermodynamic limit of the quasi-momentum distribution, 
Fig. \ref{fig:kmax_theta_hc} (b) and (c) show the system size dependence of $K_{\mathrm{max}}^{(\mathrm{a)}}$ 
and $n_{\mathrm{max}}^{L}/n_{\mathrm{max}}^{L\rightarrow \infty}$ at density $n_0 = 0.75$ in the GW approach, 
where $n_{\mathrm{max}}^{L}$ is the maximum value of the quasi-momentum distribution with the finite system size $L$ 
and $n_{\mathrm{max}}^{L\rightarrow \infty}$ corresponds to the thermodynamic limit. 
Considering the uncertainty of the quasi-momentum in the finite system $\Delta k=2\pi/L$, 
$K_{\mathrm{max}}^{(\mathrm{a)}}$ turns out to be independent of the system size. 
According to Fig. \ref{fig:nk_L120N60_hc}, the bosons condense at zero momentum 
with a sharp distribution and the fermions have a flat distribution. 
In Fig. \ref{fig:kmax_theta_hc}(c), with $\theta$ increasing from the Bose to the Fermi case, 
we find that the condensate population near the Bose side slightly depends on the system size, 
but the maximum particle number near the Fermi side is almost independent on the system size,  
since the behavior of anyons with $\theta \sim \pi$ looks much more like fermions without condensation. 
Note that we do not show here the case of bosons $\theta = 0$ because bosons condense at zero momentum. 

\section{Soft-core anyons}
\label{soft-core}

Before going to the soft-core case, we remind that within the Gutzwiller theory 
$\langle \hat{b}_i^{\dagger} \hat{b}_j \rangle $ and $\langle \hat{n}_k^{(\mathrm{b})} \rangle$ 
in Eqs. (\ref{bibj_hc1}) and (\ref{HC:nk_b_hc}) turn out to be  
independent of the fractional phase $\theta$ of hard-core anyons because $\Delta\beta_0 = 2m\pi$ 
and the local state $|2\rangle$ is not occupied, i.e. $F_2 = 0$. 
Considering the occupation-dependent hopping in the soft-core case, 
we are going to discuss the influence of $\Delta\beta_1$ and non-vanishing $F_2$ in the following. 
The respective details of minimizing the total energy are given in Sec. \ref{appen:SC} of the Appendix. 

\subsection{Boson Quasi-momentum Distribution}
\label{soft-core1}

\begin{figure}[t]
  \centering
    \hspace*{-0.4cm}\includegraphics[width=0.8\textwidth]{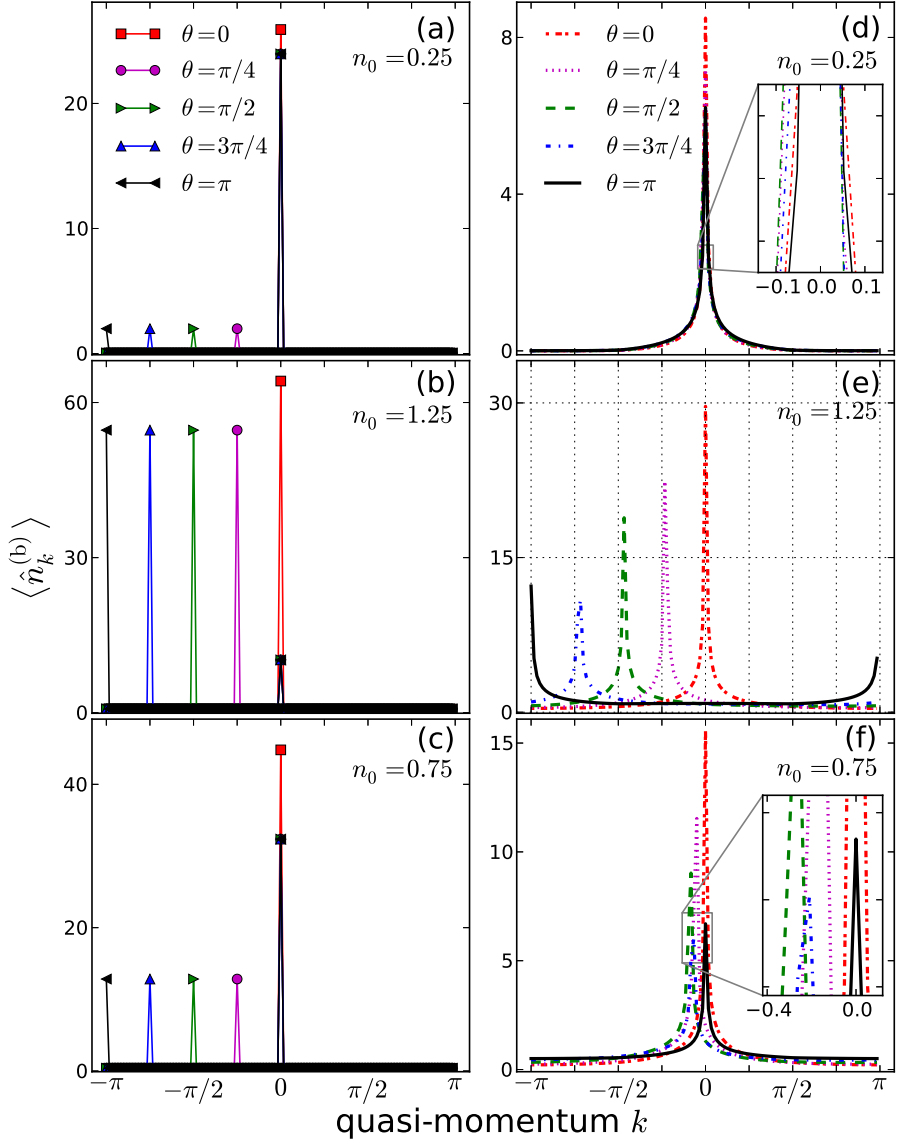}
    \caption{Quasi-momentum distribution of bosons with different statistical phase 
    $\theta$ ($J/U=0.1$, $L=120$). 
    (a), (b) and (c) from mGW at density $n_0=0.25$, $n_0=1.25$, and $n_0=0.75$. 
    (d), (e) and (f) from DMRG at density $n_0=0.25$, $n_0=1.25$, and $n_0=0.75$. 
	}
\label{fig:nk_b_L120_sf}
\end{figure}

According to Eq. (\ref{bibj}) and supposing $n_{\mathrm{max}}=2$, 
the correlation function of bosons is 
\begin{equation}\label{bibj_sf}
\langle \hat{b}_i^{\dagger} \hat{b}_j \rangle 
= \delta_{ij} n_0 + (1-\delta_{ij}) [ A + B e^{\mathrm{i}(i-j)\theta} ] \, ,
\end{equation}
where $A \equiv F_1^2 (F_0^2 + \sqrt{2}F_0 F_2)$, $B \equiv F_1^2 (\sqrt{2}F_0 F_2 + 2 F_2^2)$ and $A+B=C$. 
Using Eq. (\ref{nk_b}), the quasi-momentum distribution of bosons can be written as 
\begin{equation}\label{nk_b_sf}
\langle \hat{n}_k^{(\mathrm{b})} \rangle = n_0\! -\! C + A L \delta_{k,0} 
+ B \frac{1-\cos[(k+\theta)L]}{L[1-\cos(k+\theta)]} 
\end{equation}
and yields in the thermodynamic limit
\begin{equation}\label{nk_b_sf_inf}
\langle \hat{n}_k^{(\mathrm{b})} \rangle 
\xrightarrow[]{L \rightarrow \infty} n_0\! -\! C + A \delta(k) + B \delta(k+\theta) \, .
\end{equation}
Here, compared with the $\theta$-independent $\langle \hat{n}_k^{(\mathrm{b})} \rangle$ 
of the hard-core case in Eq. (\ref{HC:nk_b_hc}), 
the quasi-momentum distribution of bosons from mGW has a background distribution $n_0 - C$ 
and two peaks at $k=0$ and $k=-\theta$ with the intensities $A$ and $B$, respectively. 

Noticeably, if comparing the intensity $A = F_1^2 (F_0^2 + \sqrt{2}F_0 F_2)$ of the $k=0$ peak with the intensity 
$B \!=\! F_1^2 (\sqrt{2} F_0 F_2 + 2 F_2^2)$ of the $k=-\theta$ peak, we find that the discrepancy between 
the intensities $A$ and $B$ is determined by the occupation $F_0^2$ of $|0\rangle$ and $F_2^2$ of $|2\rangle$. 
As these occupations $F_0^2$ and $F_2^2$ change with the density $n_0$, the resulting quasi-momentum 
distribution of bosons reveals a characteristic density dependency. 
Therefore, we discuss now different density regimes. 

In the dilute limit $n_0 \ll 1$, the positive Hubbard $U$ depresses the state $|2\rangle$ and $F_2^2 \sim 0$, 
thus the properties of $\langle \hat{n}_k^{(\mathrm{b})} \rangle$ are similar to that of the hard-core case. 
Fig. \ref{fig:nk_b_L120_sf}(a) shows the quasi-momentum distribution of bosons at density $n_0=0.25$ 
and $J/U=0.1$ in the mGW approach. 
For $\theta=0$, we have $\langle \hat{n}_k^{(\mathrm{b})} \rangle = n_0 - C + C\delta(k)$, 
which demonstrates that there exists only one peak of the intensity $C=A+B$ at $k=0$ 
and the quasi-momentum distribution is symmetric. 
For the case of $0 < \theta \leq \pi$, we have (\ref{nk_b_sf_inf}), which demonstrates 
that there exists one much lower peak of the intensity $B$ at the negative momentum $k=-\theta$ 
than that of the intensity $A$ at $k=0$ due to $F_2^2 \ll F_0^2$. 
Here, we should note that $\langle \hat{n}_k^{(\mathrm{b})} (\theta=\pi) \rangle$, 
which is plotted for slightly asymmetric interval $-\pi \leq k < \pi$ in Fig. \ref{fig:nk_b_L120_sf}(a), 
is also a symmetric distribution since $\langle \hat{n}_k^{(\mathrm{b})} \rangle$ has period $2\pi$. 

The results of DMRG with the same parameters are shown in Fig. \ref{fig:nk_b_L120_sf}(d), 
where the symmetric quasi-momentum distribution of $\theta=0$ is a little sharper than that of $\theta=\pi$. 
But for the case of  $0 < \theta < \pi$, there exists an asymmetric quasi-momentum distribution 
and more bosons have negative momentum according to the inset of Fig. \ref{fig:nk_b_L120_sf}(d). 
In general, quantum fluctuations broaden $\langle \hat{n}_k^{(\mathrm{b})} \rangle$ from mGW. 
Taking this into account, if we would add the effect of quantum fluctuations 
on the mGW quasi-momentum distribution, the peak at $k=0$ and  $k=-\theta$ would broaden. 
Since the peak at $k=-\theta$ from mGW is pretty low, a broadening would be merged into 
the background of the quasi-momentum distribution. Therefore, at small particle densities, 
the mGW approach gives qualitatively the same quasi-momentum distribution 
as DMRG, i.e. the peaks locate at $k=0$ and more bosons have negative momentum. 

\begin{figure}[b]
  \centering
    \hspace*{-0.5cm}\includegraphics[width=0.8\textwidth]{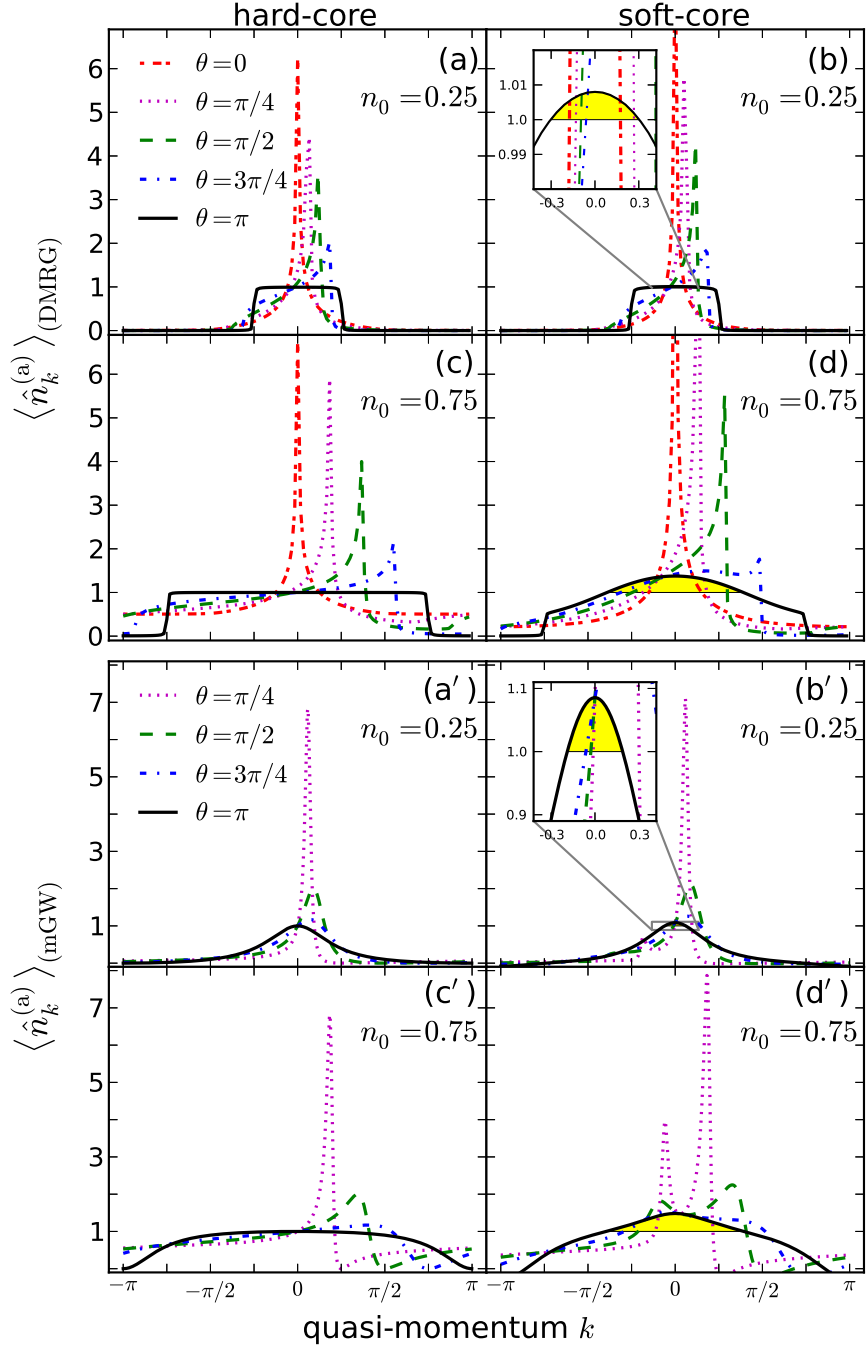}
    \caption{Quasi-momentum distribution of anyons with different statistical phase $\theta$ ($J/U=0.1$). 
    (a), (b), (c) and (d) from DMRG ($L=120$) at density $n_0=0.25$ and $n_0=0.75$. 
    (a$^\prime$), (b$^\prime$), (c$^\prime$) and (d$^\prime$) from mGW in the thermal dynamic limit 
    at density $n_0=0.25$ and $n_0=0.75$. 
    The left column (a), (c), (a$^\prime$), (c$^\prime$) and the right column (b), (d), (b$^\prime$), (d$^\prime$) 
    refer to the hard-core case and the soft-core case, respectively. 
	}
\label{fig:nk_a_hc2sf}
\end{figure}

For the high density $n_0 = 1.25$ in Fig. \ref{fig:nk_b_L120_sf}(b) and (e), 
there exists more than one particle per site on average. 
Thus the vacuum state $|0\rangle$ is almost unoccupied in the ground state, i.e. $F_0^2 \ll F_2^2$. 
For $0< \theta \leq \pi$, Fig. \ref{fig:nk_b_L120_sf}(b) clearly shows that the high peaks 
of the quasi-momentum distribution exactly appear at $k=-\theta$ in comparison with that at $k=0$. 
The results of mGW qualitatively coincide with that of DMRG in Fig. \ref{fig:nk_b_L120_sf}(d) 
although the momentum value $k$ of the DMRG peaks is a little larger than $-\theta$ 
because the peaks at $k=-\theta$ compete with the peaks at $k=0$ under the influence of quantum fluctuation. 
Of course, $\theta=0$ and $\theta=\pi$ still give the symmetric quasi-momentum distribution of bosonic operators. 

Fig. \ref{fig:nk_b_L120_sf}(c) and (f) show the quasi-momentum distribution at the density $n_0 = 0.75$ 
in the middle of $n_0 = 0.25$ and $n_0 = 1.25$. The competition between the $k=0$ peak of the intensity $A$ and 
the $k=-\theta$ of the intensity $B$ of the quasi-momentum distribution leads to final peak shifts from $k=0$ 
to the negative momentum but comes back to $k=0$ while increasing $\theta$ from $0$ to $\pi$
as shown in Fig. \ref{fig:nk_b_L120_sf}(f). 
Note that this peculiar behaviour is not reproduced by mGW in Fig. \ref{fig:nk_b_L120_sf}(c). 

If we think about the occupation-dependent factor $e^{\mathrm{i}\theta \hat{n}_{j}}$ in (\ref{bosons}) again, 
we recognize in Fig. \ref{fig:nk_b_L120_sf} a clear physical picture of the ground state: 
i) for a small density $n_0 < 1$, the states $|0\rangle$ and $|1\rangle$ are preferred and 
$\Delta \beta_0 = 2m\pi$ is the leading term, which gives a quasi-momentum peak of bosons at $k \sim 0$; 
ii) for a density $1< n_0 < 2$, the states $|1\rangle$ and $|2\rangle$ are preferred and 
$\Delta \beta_1 = 2l\pi - \theta$ is the leading term, which gives a quasi-momentum peak of bosons at $k \sim -\theta$. 
In order to deal with the density $2 < n_0$, we should truncate the mGW approach at $n_{\mathrm{max}} = 3,4,\cdots$. 
For example assuming $n_{\mathrm{max}} = 3$, the quasi-momentum distribution of bosons shows an additional peak 
of $k=-2\theta$. For the density $2 < n_0 < 3$, the leading peak of the quasi-momentum distribution of bosons 
appears at the momentum $k \sim -2\theta$. 

\subsection{Anyon Quasi-momentum Distribution}
\label{soft-core2}

According to the Eqs. (\ref{aiaj}) and (\ref{op_aver}) in the Appendix, 
the anyonic correlation function is written as 
\begin{subequations}\label{aiaj_sf}
\begin{align}
\langle \hat{a}_i^{\dagger} \hat{a}_j \rangle 
&\xrightarrow[]{i<j} [ A + B e^{\mathrm{i}(i-j+1)\theta} ] \prod_{i<l<j} w \, ,\label{aiaj1_sf} \\
\langle \hat{a}_i^{\dagger} \hat{a}_j \rangle 
&\xrightarrow[]{i>j} [ A + B e^{\mathrm{i}(i-j-1)\theta} ] \prod_{j<l<i} w^{*} \, . \label{aiaj2_sf}
\end{align}
\end{subequations}
where $w=F_0^2 + F_1^2 e^{\mathrm{i}\theta} + F_2^2 e^{\mathrm{i}2\theta} \equiv W e^{\mathrm{i}\chi}$. 
If $\theta = 0$, we have $w =F_0^2 + F_1^2 + F_2^2 = 1$, otherwise the absolute value is $W < 1$. 

In the Bose limit $\theta = 0$, the correlation function of anyons is 
\begin{equation}
\langle \hat{a}_i^{\dagger} \hat{a}_j \rangle = \delta_{ij} n_0 + (1-\delta_{ij})C \, .
\end{equation}
Using Eq. (\ref{nk_a}), the ground-state quasi-momentum distribution of anyons can be written as 
\begin{equation}\label{nk_a0_sf}
\langle \hat{n}_k^{(\mathrm{a})} \rangle = n_0 - C + C L \delta_{k,0} \, .
\end{equation}

If $\theta \neq 0$ and then $|w| < 1$, the quasi-momentum distribution of anyons is written as 
\begin{equation}\label{nk_a1_sf_L}
\langle \hat{n}_k^{(\mathrm{a})} \rangle = 
n_0\! + \frac{A}{L} \! \Bigl[e^{-\mathrm{i}k} \frac{(L\!-\!1)\!-\! Lu \!+\! u^L}{(1-u)^2} \! + \! \mathrm{c.c.} \Bigr] 
+ \frac{B}{L} \! \Bigl[e^{-\mathrm{i}k} \frac{(L\!-\!1)\!-\! Lv \!+\! v^L}{(1-v)^2}  \! + \! \mathrm{c.c.} \Bigr]\,,
\end{equation}
where $u=we^{-\mathrm{i}k}$ and $v=we^{-\mathrm{i}(k+\theta)}$. In the thermodynamic limit, this reduces to 
\begin{equation}\label{nk_a1_sf}
\langle \hat{n}_k^{(\mathrm{a})} \rangle \xrightarrow[|w|<1]{L\rightarrow \infty} 
n_0\! + \! A \frac{2\cos k -2W\cos \chi }{1-2W\cos(k\! -\! \chi) \! + \! W^2} 
+ B \frac{2\cos k -2W\cos(\theta - \chi)}{1-2W\cos[(k+\theta)\! -\! \chi] \! + \! W^2} \,,
\end{equation}
where $W=|w|$ and $\chi = \mathrm{arg}(w)$. 

In the hard-core limit $F_2^2 = 0$, so we have $w=z$, $\chi = \varphi$, 
$A=n_0(1-n_0)$, and $B = 0$, thus Eq. (\ref{nk_a1_sf}) reduces to Eq. (\ref{nk_a_hc2}).  
For small densities $n_0 \ll 1$, which implies $F_2^2 \sim 0$, 
we expect the quasi-momentum distribution of anyons 
in the soft-core case to be similar to that in the hard-core case. 
This can be shown by comparing Fig. \ref{fig:nk_a_hc2sf} (a) and (b) from DMRG 
or (a$^\prime$) and (b$^\prime$) from mGW at the density $n_0=0.25$. However, for the fractional phase 
$\theta = \pi$ in the soft-case, we also note that there exists a quite small area 
with the property $\langle \hat{n}_k^{(\mathrm{a})} \rangle > 1$ that is shown as the shaded (yellow in color) 
zone in the insets of Fig. \ref{fig:nk_a_hc2sf} (b) and (b$^\prime$). 
The shaded area increases while increasing the density, for example $n_0 = 0.75$ is shown 
in Fig. \ref{fig:nk_a_hc2sf} (d) and (d$^\prime$). 
This result reveals that anyons with the statistical phase $\theta = \pi$ are pseudofermions, i.e. 
possibly more than one identical anyon exist in one state but they behave like a fermion 
while exchanging two anyons on different sites. Comparing the results from DMRG and mGW 
in Fig. \ref{fig:nk_a_hc2sf}, we demonstrate again that the mGW approach grasps 
the main features of the quasi-momentum distribution of anyons. 
Finally, we should note that the hard-core case is more suitable to describe the step-like behaviour of 
anyons ($\theta = \pi$) than the soft-core case. 

\begin{figure}[t]
  \centering
   \includegraphics[width=0.6\textwidth]{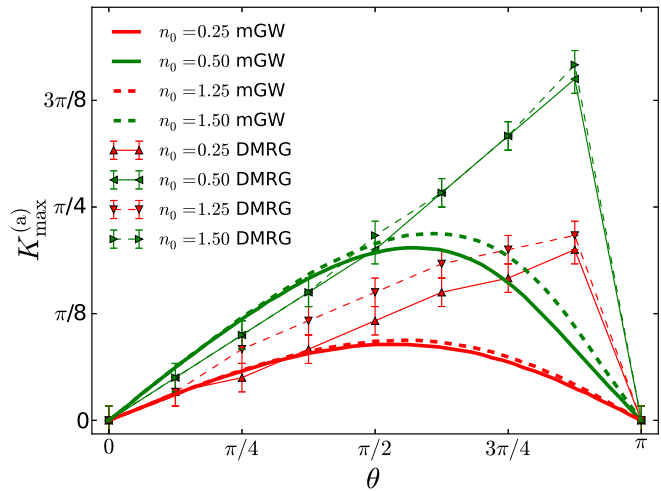}
   \caption{Quasi-momentum $K_{\mathrm{max}}^{(\mathrm{a)}}$ 
   plotted against the fractional phase $\theta$ of anyons 
   resulting from DMRG ($L=120$) and mGW (in the thermodynamic limit $L \rightarrow \infty$) 
   with parameters $J/U=0.1$, density $n_0 = 0.25, 0.5, 1.25$, and $1.5$, respectively. 
   The uncertainty of the quasi-momentum is of the order $\Delta k = 2\pi /L$. 
   }
\label{fig:kmax_theta_sf}
\end{figure}

Fig. \ref{fig:kmax_theta_sf} shows the fractional phase $\theta$-dependence of anyonic 
$K_{\mathrm{max}}^{(\mathrm{a)}}$ at the densities $n_0 = 0.25$, $0.5$, $1.25$, and $1.5$. 
In general, for the density $0 < n_0 < 1$ and $0 < \theta < \pi$, $K_{\mathrm{max}}^{(\mathrm{a)}}$ at density 
$n_0 = 0.5$ is larger than that at $n_0 = 0.25$, which coincides with the finding of the hard-core case. 
Moreover, for $1 < n_0 < 2$, the DMRG results show that the peak position $K_{\mathrm{max}}^{(\mathrm{a)}}$ 
of the quasi-momentum distribution at density $n_0 = 1.25$ coincides with that at density $n_0 = 0.25$ 
within uncertainty. 
This is also shown by the DMRG results at density $n_0 = 1.5$ and $n_0 = 0.5$. 
However, the mGW approach in the thermodynamic limit shows that $K_{\mathrm{max}}^{(\mathrm{a)}}$ 
at density $n_0+1$ is a little larger than that at density $n_0$, especially for the zone $\theta \sim \pi$. 

\section{Conclusions}
\label{Conclusions}

In summary, we have studied the ground-state property of the 1D Anyon-Hubbard model. 
With the help of a fractional version of the Jordan-Wigner transformation, 
the Anyon-Hubbard model is mapped to the occupation-dependent hopping Bose-Hubbard model 
and, thus, the Hilbert space of anyons can be constructed from that of bosons. 
By calculating the two-point correlation function of creation and annihilation operators 
of bosons and anyons, we investigate the quasi-momentum distributions 
interpolating between Bose-Einstein statistics and Fermi-Dirac statistics. 
Theoretically, in order to include the influence of the fractional phase of anyons on the many-body wavefunction, 
we modify the classical Gutzwiller mean-field approach and get an analytic expression for 
the quasi-momentum distributions of anyons and bosons. 
In order to test the accuracy of the mGW mean-field approach, we use DMRG for numerical calculations. 

In the hard-core case, the results show that the bosons condensate at zero momentum and have 
a symmetric quasi-momentum distribution around zero momentum. Due to the nonlocal string property, 
more anyons are shifted to a positive momentum and have an asymmetric quasi-momentum distribution, 
where the peak position depends on both the fractional phase and the particle number density. 

In the soft-core case, the results show that the quasi-momentum peaks of bosons strongly depend on 
the particle number density. At density $n_0 < 1$, $\Delta \beta_0 = 2m\pi$ leads to 
a quasi-momentum peak of bosons at $k \sim 0$. At density $1< n_0 < 2$, 
$\Delta \beta_1 = 2l\pi - \theta$ leads a quasi-momentum peak of bosons at $k \sim -\theta$. 
Again, the quasi-momentum distribution of anyons show nonlocal string behaviour and yield 
similar features as in the hard-core case. Furthermore, anyons with $\theta=\pi$ are pseudofermions, i.e., 
there exists more than one identical anyon in one state in the soft-core case. 
However, in the hard-core case, the quasi-momentum distribution of anyons is a typical step-like function. 

\section*{Acknowledgments}

The authors would like to thank B. Abdullaev, A. Eckardt, S. Hu, D. Morath, M. Roncaglia, I. Schneider, 
S. Srivastava, C. Str\"ater, D. Strassel, T. Wang, and X.-F. Zhang for valuable discussions. 
This work was supported by Science Foundation of Heilongjiang (A201002), the Chinese Scholarship Council (CSC), 
the German Research Foundation (DFG) via the Collaborative Research Center SFB/TR49, and 
the Volkswagen Foundation via the German-Uzbek Project \textit{Anyon Physics of Ultracold Atomic Gases}.

\appendix
\section{Gutzwiller Mean-Field Approach}
\label{appen:GW}

Based on the Gutzwiller wave function (\ref{Awf_GW}), the interaction energy turns out to be 
\begin{equation}\label{E_int}
\begin{split}
\mathcal{E}_{\mathrm{int}} & = \sum_{j} \langle \Phi_j | \frac{U}{2} \hat{n}_j(\hat{n}_j-1) | \Phi_j \rangle \\
                  & = \frac{U}{2} \sum_{j} \sum_{n=0}^{n_{\mathrm{max}}} |f_n^{(j)}|^2 n(n-1) \, ,
\end{split}
\end{equation}
whereas the chemical potential term within a grand-canonical description reads 
\begin{equation}\label{E_mu}
\mathcal{E}_{\mu} = \sum_{j} \langle \Phi_j | (-\mu) \hat{n}_j | \Phi_j \rangle 
                  = -\mu  \sum_{j} \sum_{n=0}^{n_{\mathrm{max}}} |f_n^{(j)}|^2 n \, .
\end{equation}
In the same way, the expectation value of the kinetic energy term can be expressed as 
\begin{equation}\label{E_kin}
\begin{split}
\hspace*{-0.3cm}\mathcal{E}_{\mathrm{kin}} 
   \!\! &=\!\! -J\!\sum_{j}\!(\!\langle \Phi_j |\hat{b}_{j}^{\dagger} e^{\mathrm{i}\theta\hat{n}_j}|\Phi_j\rangle 
                    \langle \Phi_{j+1} | \hat{b}_{j+1} | \Phi_{j+1} \rangle \!\! + \! \mathrm{c.c.}  \! ) \\
   \!\! &= \!\! -J\!\sum_{j}\!( \langle \hat{b}_{j}^{\dagger} e^{\mathrm{i}\theta\hat{n}_j} \rangle 
                    \langle \hat{b}_{j+1} \rangle  + \mathrm{c.c.} ) \, ,
\end{split}
\end{equation}
where the expectation values $\langle \hat{O}_j \rangle \equiv \langle \Phi_j | \hat{O}_j | \Phi_j \rangle$ yield 
\begin{subequations}\label{op_aver}
\begin{align}
\langle \hat{b}_j \rangle 
  & = \sum_{n=0}^{n_{\mathrm{max}}} f_n^{(j)*}  f_{n+1}^{(j)} \sqrt{n+1} 
    = \langle \hat{b}_j^{\dagger} \rangle ^{*} \, , \label{b_aver}\\
\langle \hat{b}_{j}^{\dagger} e^{\mathrm{i}\theta\hat{n}_j} \rangle \!\!
  & = \!\! \sum_{n=0}^{n_{\mathrm{max}}} f_n^{(j)}  f_{n+1}^{(j)*} \sqrt{n+1} e^{\mathrm{i}\theta n} \!\!
    = \!\! \langle e^{-\mathrm{i}\theta\hat{n}_j} \hat{b}_{j} \rangle ^{*} \, , \label{bn_aver} \\
\langle \hat{n}_j \rangle 
  & = \sum_{n=0}^{n_{\mathrm{max}}} |f_n^{(j)}|^2 n \, , \label{n_aver}\\
\langle e^{\pm \mathrm{i}\theta\hat{n}_{j}} \rangle
  & = \sum_{n=0}^{n_{\mathrm{max}}} |f_n^{(j)}|^2 e^{\pm \mathrm{i}\theta n} \, . \label{en_aver}
\end{align}
\end{subequations}

Using Eq. (\ref{op_aver}), the Gutzwiller kinetic energy reduces to 
\begin{equation}\label{E_kin_f}
\mathcal{E}_{\mathrm{kin}} =  -J \sum_{j}  \sum_{m,n}^{n_{\mathrm{max}}} \sqrt{(n+1)(m+1)} \bigl[ f_n^{(j)} f_{n+1}^{(j)*} f_m^{(j+1)*} f_{m+1}^{(j+1)} e^{\mathrm{i}\theta n} \! + \! \mathrm{c.c.} \bigr]   \, .
\end{equation}
Using the polar decomposition $f_n^{(j)} = F_n^{(j)} e^{ \mathrm{i}\alpha_n^{(j)} } $, we have 
\begin{equation}\label{f_polar}
f_n^{(j)} f_{n+1}^{(j)*} f_m^{(j+1)*} f_{m+1}^{(j+1)} = 
F_n^{(j)} F_{n+1}^{(j)} F_m^{(j+1)} F_{m+1}^{(j+1)} 
e^{ \mathrm{i} [ \beta_n^{(j)} - \beta_{m}^{(j+1)} ] } \,,
\end{equation}
where $\beta_n^{(j)} \equiv \alpha_n^{(j)} - \alpha_{n+1}^{(j)} $ denotes the difference 
between the phase of the probability amplitude $f_n^{(j)}$ of finding $n$ bosons and 
$f_{n+1}^{(j)}$ of finding $n+1$ bosons on site $j$. 

The ground state of the system with given parameters is determined from varying the coefficients $f_n^{(j)}$ 
and minimizing the total energy. This can either be done for a given chemical potential $\mu$, or for 
a given mean particle number $n_0 = \langle \hat{n} \rangle = \sum_{n=0}^{n_{\mathrm{max}}} n F_n^2$ 
according to Eq. (\ref{n_aver}). In this paper, we assume that $n_0$ is given, so the chemical potential 
$\mathcal{E}_{\mu} = -\mu n_0 L$ resulting from (\ref{E_mu}) does not depend on $f_n^{(j)}$. 
Thus, after having determined the Gutzwiller coefficients $f_n^{(j)}$, 
the chemical potential $\mu$ is fixed by 
\begin{equation}
\mu = \frac{1}{L} \frac{\partial (\mathcal{E}_{\mathrm{int}} + \mathcal{E}_{\mathrm{kin}})}{\partial n_0} \,.
\end{equation}
\\[1ex]

\subsection{Hard-Core Case}
\label{appen:HC} 
In the hard-core limit, bosons are impenetrable and each site contains at most one particle, i.e., $n_{\mathrm{max}}=1$. 
In the GW approach, the normalization condition (\ref{normalization}) reduces to $F_0^2+F_1^2=1$. 
For a given particle number $N$, the mean density of particle number is given by $n_0 = N/L = F_1^2$ according to 
the Eq. (\ref{n_aver}). Thus the absolute values of both Gutzwiller amplitudes are determined according to 
\begin{equation}\label{normalization_hd}
F_0^2=1-n_0, \quad F_1^2=n_0 \, .
\end{equation}
The expectation value of the energy per lattice site $E_{\mathrm{tot}}=\mathcal{E}_{\mathrm{tot}}/L$ is then given by 
\begin{equation}
\begin{split}
E_{\mathrm{tot}}^{\mathrm{cGW}} &= -\mu n_0 - 2J n_0 (1-n_0) \, , \\
E_{\mathrm{tot}}^{\mathrm{mGW}} &= -\mu n_0 - 2J n_0 (1-n_0) \cos\Delta\beta_0 \, ,
\end{split}
\end{equation}
where $E_{\mathrm{tot}}^{\mathrm{cGW}}$ and $E_{\mathrm{tot}}^{\mathrm{mGW}}$ 
refer to the cGW and mGW approach, respectively. 
The ground state is determined by minimizing the energy $E_{\mathrm{tot}}^{\mathrm{cGW}}$ 
and $E_{\mathrm{tot}}^{\mathrm{mGW}}$. This leads to 
\begin{equation}\label{dbeta_eq0_hc}
\Delta\beta_0 = 2m\pi \, , 
\end{equation}
where $m=0,\pm 1, \pm 2, \cdots$. This expression shows that the mGW approach reduces to the cGW approach 
for hard-core anyons. 

\subsection{Soft-Core Case}
\label{appen:SC}

Using the normalization condition (\ref{normalization}) and Eq. (\ref{n_aver}) in the GW approach, we have 
\begin{equation}\label{F_value_sf}
\begin{split}
F_0^2 + F_1^2 + F_2^2 &= 1 \, , \\
F_1^2 + 2 F_2^2 &= n_0 \, ,
\end{split}
\end{equation}
where we suppose $n_{\mathrm{max}} = 2$ as an approximation. 
The expectation value of the energy per lattice site reads now
\begin{equation}\label{etot0_sf}
\begin{split}
\hspace*{-0.35cm} E_{\mathrm{tot}}^{\mathrm{mGW}} &\!\!=\!\! -\mu n_0 + U F_2^2 -2J F_1^2  \Bigl\{ F_0^2\cos(\Delta\beta_0) + 2 F_2^2 \cos(\Delta\beta_1+\theta) \\
&\mspace{9.0mu} + 2\sqrt{2}F_0 F_2 \cos \Bigl(\frac{\Delta\beta_0\!+\!\Delta\beta_1\!+\!\theta}{2} \Bigr) 
              \cos \Bigl(\beta_0\!-\!\beta_1\!-\!\frac{\theta}{2} \Bigr)  \Bigr\} \,.
\end{split}
\end{equation}
In order to minimize (\ref{etot0_sf}), we must choose for the respective phases 
\begin{equation}\label{theta_sf}
\begin{split}
\Delta\beta_0 = 2 m \pi \, , &\quad \Delta\beta_1 + \theta = 2 l \pi \, , \\
\beta_0-\beta_1-\frac{\theta}{2} &= (m+l) \pi \, ,
\end{split}
\end{equation}
where $m,l = 0,\pm 1,\pm 2, \cdots $. 
This demonstrates that the non-vanishing $\theta$ influences the phase of the many-body wavefunction. 
Taking into account Eq. (\ref{F_value_sf}), the ground state can be found by varying the Gutzwiller coefficients 
$F_0, F_1, F_2$ and minimizing the energy 
\begin{equation}\label{etot1_sf}
E_{\mathrm{tot}}^{\mathrm{mGW}} = -\mu n_0 + U F_2^2 -2J C \, ,
\end{equation}
where $C\equiv F_1^2 (F_0^2 + 2\sqrt{2}F_0 F_2 + 2 F_2^2)$. Note that the resulting occupations $F_0^2$, $F_1^2$, 
and $F_2^2$ of the states $|0\rangle$, $|1\rangle$, and $|2\rangle$ are not affected by $\theta$. 
Obviously, in the hard-core limit $J/U \rightarrow 0$, we reproduce $F_2 \rightarrow 0$. 
Thus, in that case Eq. (\ref{F_value_sf}) goes over to (\ref{normalization_hd}), 
so the calculation reproduces the hard-core case of the last section. 

Furthermore, if we set $\Delta\beta_0 = 2m\pi$ and $\Delta\beta_1 = 2l\pi$ in Eqs. (\ref{etot0_sf}), 
the mGW approach reduces to the cGW approach, for which the energy is 
\begin{equation}\label{etot0_cGW_sf}
\begin{split}
\hspace*{-0.2cm} E_{\mathrm{tot}}^{\mathrm{cGW}} &= -\mu n_0 \!+\! U F_2^2 \!-\! 
                  2J F_1^2  \Bigl\{ F_0^2 + 2 F_2^2 \cos(\theta) \\
&\quad + 2\sqrt{2}F_0 F_2 \cos \Bigl( \frac{\theta}{2} \Bigr) 
                          \cos \Bigl( \beta_0-\beta_1-\frac{\theta}{2} \Bigr) \Bigr\} \,.
\end{split}
\end{equation}
Comparing Eq. (\ref{etot1_sf}) with (\ref{etot0_cGW_sf}), 
it is straightforward to conclude that $E_{\mathrm{tot}}^{\mathrm{mGW}} \leq E_{\mathrm{tot}}^{\mathrm{cGW}}$, 
which shows that the mGW approach is superior to the cGW approach. 
Therefore, we restrict ourselves to evaluate the mGW approach for the soft-core case.

\section*{References}

\providecommand{\newblock}{}

\end{document}